\documentclass[%
 reprint,
 amsmath,amssymb,
 aps,
]{revtex4-1}

\usepackage{amsmath}    
\usepackage{graphicx}   
\usepackage{docs}%
\usepackage{bm}%
\usepackage[colorlinks=true,linkcolor=blue]{hyperref}%
\expandafter\ifx\csname package@font\endcsname\relax\else
 \expandafter\expandafter
 \expandafter\usepackage
 \expandafter\expandafter
 \expandafter{\csname package@font\endcsname}%
\fi
\newcommand{\eq}[2]{ \begin{equation} \label{#1}\begin{split} #2 \end{split} \end{equation} }

\def\av#1{\langle{#1}\rangle}
\def\pti#1{#1}
\newcommand{\qq}{\,,\qquad}
\def\be#1{\begin{equation}\label{#1}}
\def\ee{\end{equation}}
\def\ba#1#2{\be{#1}\begin{array}{#2}}
\def\ea{\end{array}\ee}
\def\={\stackrel{\mbox{\scriptsize def}}{=}}

\begin{document}

\title{Thermal echo in a finite one-dimensional harmonic crystal}
\author{A.S. Murachev}
\email{andrey.murachev@gmail.com}
\affiliation{Department of Theoretical Mechanics, Peter the Great St.~Petersburg Polytechnic University}

\author{A.M. Krivtsov}
\email{akrivtsov@bk.ru}
\affiliation{Department of Theoretical Mechanics, Peter the Great St.~Petersburg Polytechnic University}
\author{D.V. Tsvetkov}
\email{dvtsvetkov@ya.ru }
\affiliation{Department of Theoretical Mechanics, Peter the Great St.~Petersburg Polytechnic University}%

\date{02 July 2018}

\begin{abstract}
An instant homogeneous thermal perturbation in the finite harmonic one-dimensional crystal is studied. Previously it was shown that for the same problem in the infinite crystal the kinetic temperature oscillates with decreasing amplitude described by the Bessel function of the first kind. In the present paper it is shown that in the finite crystal
this behavior is observed only until a certain period of time when a sharp increase of the oscillations amplitude is realized. This phenomenon, further refereed to as the thermal echo, is realized periodically, with the period proportional to the crystal size. The amplitude for each subsequent echo is lower than for the previous one. It is obtained analytically  that the time-dependance of the kinetic temperature can be described by an infinite sum of the Bessel functions with multiple indexes. It is also shown that the thermal echo in the thermodynamic limit is described by the Airy function.
\end{abstract}
\maketitle

\section{Introduction}
Analytical and experimental results demonstrate an anomalous nature of thermal processes in ultrapure materials~\cite{Rieder1967,Gendelman,Krivtsov 2015 DAN, Kuzkriv2016,Lepri2003,Chang2007,GuzDmi17,Bonetto2000,Hsiao15}.
These processes can be caused by shock waves~\cite{Kanel1996,Holian1980,Holian2010,Hoover2014} or by ultrashort laser pulses~\cite{Silva2015,јшитков2013,Inogamov2012,Poletkin 2012,bykov2014,Indeitsev2009,Indeitsev2016}.
One-dimensional crystal lattices are subject of an extensive research since they admit analytical solutions and
allow verification of fundamental phenomena inherent also
for higher dimensional systems~\cite{psyRev2017, GuzDmi17-2,Godelman10,Babenkov2016,Shishkina16}.
An analytical description of the thermal processes in non-equilibrium harmonic crystals can be obtained on the basis of the covariance analysis~\cite{Rieder1967,Krivtsov 2015 DAN,Kuzkriv2016,Krivtsov 2014 DAN}.
The corresponding description of the anomalous heat propagation is presented in papers~\cite{Krivtsov 2015 DAN, Krivtsov 2015 arXiv, Gavrilov2018} for one dimension and in works~\cite{Kuzkriv2016, Kuzkriv2017, KuzArxiv2017, TsaKuz2018} for two and three dimensions.

One of the specific phenomena of the nonequilibrium thermal processes in discrete molecular systems are the high-frequency oscillations of the kinetic temperature, which have long been known from the results of numerical simulation \cite{Allen1987}. Covariance analysis admit analytical description of this phenomenon, for example in the case of the instantaneous heat perturbation of the simplest model of one-dimensional harmonic crystal these oscillations  are described by the Bessel function of the first kind and the zero order~\cite{Krivtsov 2014 DAN}, similar result was obtained earlier by direct analytical solution of the equations of the atoms motion by Ilya~Prigogine~\cite{Prig}. For the case of the one-dimensional crystal on an elastic substrate the same problem is solved in~\cite{Babenkov2016}, for higher dimensions in papers~\cite{Kuzkriv2016, KuzArxiv2017}.

Unlike previous papers~\cite{Krivtsov 2015 DAN,Krivtsov 2014 DAN,Babenkov2016,Krivtsov 2015 arXiv}, where the main attention is focused on infinite crystals, the present paper investigates thermal processes in \textit{finite} crystals~\cite{CheS2017}.
The systems with a finite number of particles is of practical importance, especially because nanotechnologies are actively developing \cite{Goldstein2007,Golovnev2006,Korobeynikov2012,KrivMoroz2002,Ўумска€2017}.
In the current paper it is demonstrated that
for the finite crystals the oscillation amplitude is decreasing only until a certain moment of time when a sharp increase of the amplitude of the kinetic temperature oscillations is realized. This phenomenon can be interpreted as a thermal echo, which will be analyzed in details in the presented paper.
Exact and asymptotic formulas describing the oscillations of the kinetic temperature will be obtained. These results, in particular, are important for description of the anomalous heat propagation in ultrapure materials~\cite{Krivtsov 2015 DAN,Lepri2003,Dhar2014}.

\section{Dynamics of the crystal}
 \subsection{The mathematical model}
We consider a one-dimensional harmonic crystal containing $N$ particles connected by harmonic springs.
The equation of motion for the particles is
\begin{equation}\label{cristallMotion}
    \ddot{u}_k
    ={\cal L}u_k \qq
    {\cal L}u_k\=\omega_e^2(u_{k+1}-2u_k+u_{k-1}),
\end{equation}
where $u_k$ is the displacement of the $k^{\rm th}$ particle \mbox{($k=1..N$)}, ${\cal L}$ is the liner deference operator, \mbox{$\omega_e=\sqrt{C/m}$} is the elementary frequency, $C$ is the stiffness of the interparticle bond, $m$ is the particle mass.
Periodic boundary conditions are used:
\begin{equation}\label{BoundCondition}
  u_0\= u_{N}\qq u_{N+1}\= u_1 .
\end{equation}
The initial conditions are
\begin{equation}\label{initialCondition}
u_k=0
\qq
v_k\=\dot{u}_k=\sigma\rho_k,
\end{equation}
where
$\rho_k$ are independent random numbers with zero mathematical expectation and unit variance, $\sigma$ is the initial velocity variance. These initial conditions correspond to an instantaneous temperature perturbation, such as perturbations caused by an ultrashort laser pulse~\cite{Silva2015,јшитков2013,Poletkin 2012}.

The initial problem (\ref{cristallMotion})--\eqref{initialCondition} describes stochastic dynamics of the particles in the crystal.
Further the deterministic equations for the statistical characteristics of motion (covariances) are analyzed to describe the thermal processes in the crystal.

\subsection{The dynamics of covariances}
One of the key statistical characteristics of the crystal is the kinetic tem\-pe\-ra\-tu\-re~$T$, which can be determined by the mathematical expectation~$\av{...}$ of the square of the centered velocity.
The corresponding formula for a one-dimensional case is
\begin{equation}\label{Temperature1}
  T=\frac{m}{k_B}\av{\tilde{v}_k^2},
\end{equation}
where
\begin{equation}\label{d0}
\tilde{v}_k\= v_k-\bar{v}\qq
\bar{v}\=\frac{1}{N}\sum_{k=0}^{N-1}v_k.
\end{equation}
Here $\bar{v}$ is the center of mass velocity and $k_B$ is the Boltzmann constant.
In order to obtain a closed system of equations for statistical characteristics it is imperative that we consider the covariances of the particle velocities~\cite{Krivtsov 2015 DAN,Krivtsov 2014 DAN,KuzArxiv2017}:
\begin{equation}\label{kappa}
  \kappa_n=\av{\tilde{v}_k\tilde{v}_{k+n}},
\end{equation}
which are the quantities characterizing motion of the particle pairs.
The following initial problem for the velocity covariancies can be obtained by differentiating the covariances using relations \eqref{cristallMotion}--\eqref{initialCondition}~(see Appendix~\ref{app00}):
\begin{equation}\label{DDOTkappa}
\begin{split}
    &\ddot\kappa_n-4{\cal L}\kappa_n=-2{\cal L}\kappa_n^0, \\
     t=0:&\quad \kappa_n=\kappa_n^0\=\sigma^2\delta_n^N-\frac{\sigma^2}{N}\qq \dot \kappa_n=0,
\end{split}
\end{equation}
where $\delta_n^N$ is the periodical Kronecker symbol: $\delta_n^N=1$ for $n$ divisible by~$N$ (including $n=0$), otherwise $\delta_n=0$.
Additionally the covariances satisfy the periodicity conditions: $\kappa_{n+N}=\kappa_n$.
After solving the initial problem~\eqref{DDOTkappa}, the kinetic temperature of the crystal can be found using formula~\eqref{Temperature1}, which can be written as
\begin{equation}\label{TempFromKap}
  T=\left.\frac{m}{k_B}\,\kappa_n\right|_{n=0}.
\end{equation}
\subsection{Representation via Bessel functions}

Solution of the initial problem~\eqref{DDOTkappa} yields the following expression for the temperature of the crystal (see Appendix~\ref{app1})
\begin{equation}\label{tNsieries}
T=\frac{\Delta T}{2}\left(1+\frac{1}{N}\sum^{N-1}_{k=1}
\cos\left(4\omega_et\sin\frac{\pi k}{N}\right)\right),
\end{equation}
where
\begin{equation}\label{DeltaT}
\Delta T=\left.\frac{m}{k_B}\,\kappa_n^0\right|_{n=0}
\end{equation}
is the temperature jump initially caused by the thermal perturbation.
This expression accurately describes the time dependence of the kinetic temperature of the harmonic crystal after an instant thermal perturbation.
Formula~(\ref{tNsieries}) can be effectively used for computations, however it is less appropriate for an analytical analysis. Therefore an alternative representation for the crystal temperature in terms of the Bessel functions is be obtained below.

Let us consider identity \cite{Abramowitz}
 \eq{linecos}{
  \cos(z\sin\vartheta)=\sum_{p=-\infty}^{\infty}J_{2p}(z)\cos(2p\,\vartheta),
 }
where $J_{2p}(t)$ is the Bessel function of the first kind of order $2p$.
Substitution~\mbox{$\vartheta={\pi k}/{N}$} and summation over $k$ yields
\eq{pre-main}{
 \frac{1}{N}\sum_{k=0}^{N-1}\cos\left(z\sin\frac{\pi k}{N}\right)
 =\sum_{p=-\infty}^{\infty}J_{2p}(z)\delta_p^N,
}
where
\begin{equation}\label{b1}
  \delta_p^N=\frac{1}{N}\sum_{k=0}^{N-1}\cos\left(p\, \frac{2 \pi k}{N}\right).
\end{equation}
As mentioned before, $\delta_p^N=1$ if $p$ is divisible by $N$, overwise ${\delta_p^N = 0}$. Identity~(\ref{b1}) is derived in appendix~\ref{app2}.
Using properties of $\delta_p^N$ formula (\ref{pre-main}) can be reduced~to
 \begin{equation}\label{sumcos}
  \begin{split}
 \frac{1}{N}\sum_{k=0}^{N-1}\cos\left(z\sin\frac{\pi k}{N}\right)=
 \sum_{p=-\infty}^{\infty}J_{2pN}(z).
 \end{split}
 \end{equation}
Substitution of the obtained formula to expression (\ref{tNsieries})
gives
\begin{equation}\label{result-a}
T=T_E+\frac{\Delta T}{2}\sum_{p=-\infty}^{\infty}J_{2pN}(4\omega_et)
,\end{equation}
where
\begin{equation}\label{result1}
T_E \= \frac{\Delta T}{2}\left(1-\frac1N\right)
\end{equation}
is an equilibrium temperature~\footnote{For an infinite crystal the kinetic temperature is tending to $T_E$; for a finite crystal the temperature is oscillating in the vicinity of $T_E$.}.
Using identity \mbox{$J_{-\mu}\equiv J_{\mu}$} formula (\ref{result-a}) can be rewritten as
\begin{equation}\label{result}
T=T_E+\frac{\Delta T}{2}\,J_0(4\omega_et)+\Delta T\sum_{p=1}^{\infty}J_{2pN}(4\omega_et).
\end{equation}
Thus, the kinetic temperature can be represented as the equilibrium temperature plus the sum of terms proportional to the Bessel functions of multiple orders.

\begin{figure}[h]
\center{\includegraphics[width=1\linewidth]{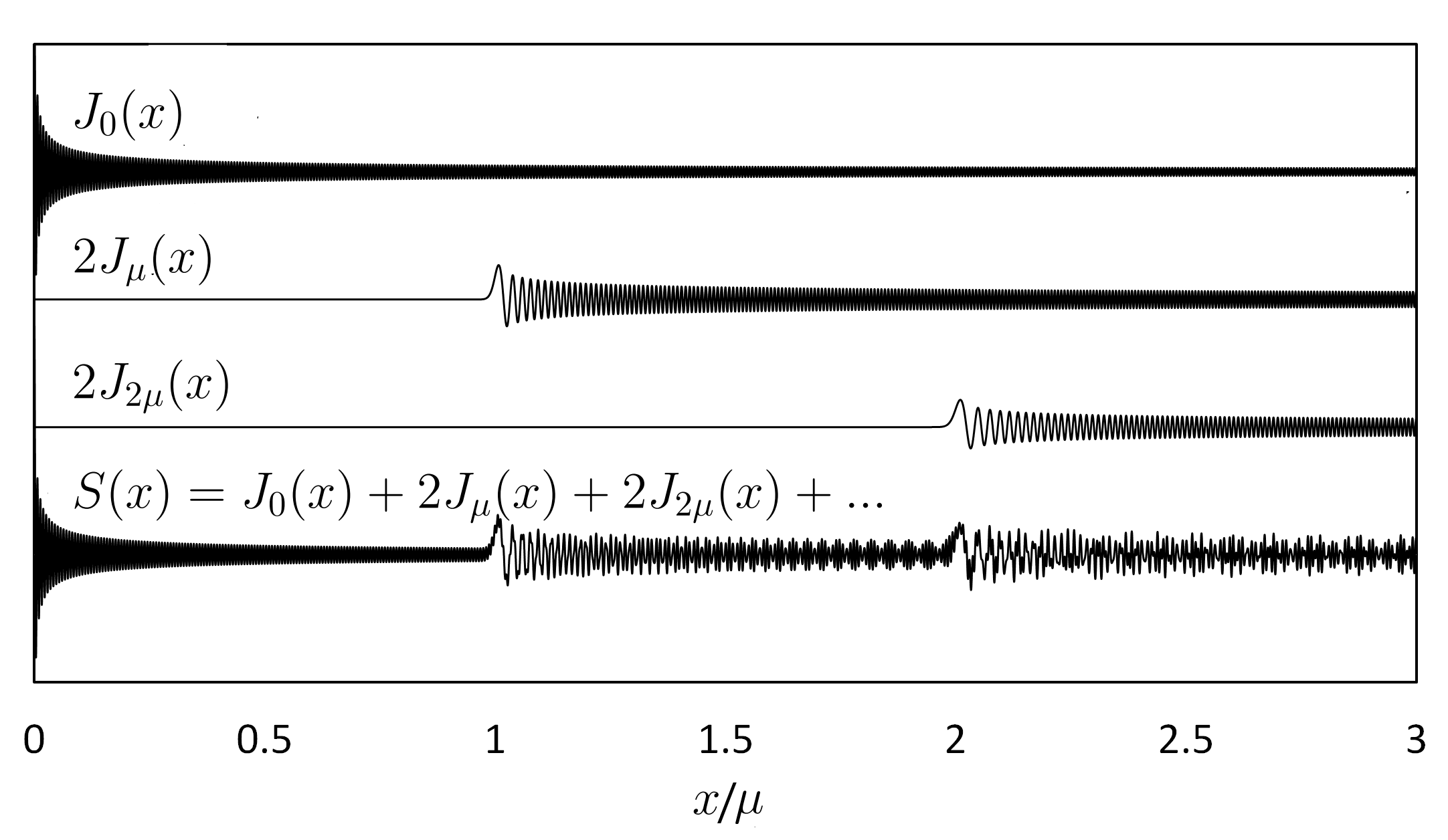}}
\caption{Bessel functions with indices multiple of $\mu=10^3$ (the top three graphs); sum of the Bessel functions (the lower graph).}
\label{ris:image0}
\end{figure}

Both expressions \eqref{tNsieries} and \eqref{result} are accurate, but expression (\ref{result}) is more suitable for the analytical analysis.
Indeed, for a positive integer index~$\mu$ the Bessel function $J_\mu(x)$ is almost zero for all positive $x$ up to a vicinity of the point~$x=\mu$.
Therefore, for a finite moment in time only a finite number of terms give noticeable contribution to representation (\ref{result}).
For illustration let us consider series
\begin{equation}\label{BesselSum}
  S(x)=J_0(x)+2J_{\mu}(x)+2J_{2\mu}(x)+...+2J_{p\mu}(x)+...
\end{equation}
Graphs for $S(x)$ and the first three terms of its representation (\ref{BesselSum}) are shown in
Fig.~\ref{ris:image0}. The figure shows that for almost~\footnote{Except a small vicinity of the right boundary $x=\mu$, where the width of this vicinity can be determined from expression \eqref{deltax}.} the entire interval~$[0,\,\mu)$ the sum $S(x)$ is determined only by the first term in expression (\ref{BesselSum}). Similarly, for almost the entire interval~$[\mu, \, 2\mu)$ the sum $S(x)$ is determined by the first two terms, and so on.

If the crystal has a nonzero initial temperature $T_0$ before the heat perturbation then the equilibrium temperature \eqref{result1} contains additional term~$T_0$:
\begin{equation}\label{result2}
T_E = T_0+\frac{\Delta T}{2}\left(1-\frac1N\right)
,\end{equation}
while expressions \eqref{result-a}, \eqref{result} remain unchanged.
This is a direct consequence of the superposition principle, which is valid for harmonic systems.

\begin{equation}\label{resultWithTe}
\begin{split}
  &\tilde{T}=T-T_E=\\
  &\frac{\Delta T}{2}\left(J_0(4\omega_et)+2\sum_{p=1}^{\infty} J_{2pN}(4\omega_et)-\frac{1}{N}
  \right),
\end{split}
  \end{equation}
where $\Delta T$ is the temperature jump~\footnote{The temperature jump is the difference between the temperature of the crystal at the point in time immediately after the temperature perturbation, and the temperature at the point in time immediately preceding the temperature perturbation.}, is caused by temperature perturbation and
$T_E\= T_0+\Delta T/2$ is the equilibrium temperature. As follows, from expression~\eqref{resultWithTe} the equilibrium temperature is the temperature reached when $ N \to \infty $, $t \to \infty $ (equilibrium state of an infinite crystal).

\section{Oscillations of the kinetic temperature}
\subsection{Thermal echo}

According to the virial theorem~\cite{Dau,Hoower2015}
in harmonic systems a time average for both kinetic and potential energies tends to the same value. Earlier studies~\cite{Krivtsov 2014 DAN, Prig} have shown that in the one-dimensional infinite harmonic crystal this equilibration process is accompanied by the energy (and, consequently, the temperature) oscillations described by the Bessel function of the zero order.
According to formula~\eqref{result-a} or~\eqref{result} the same process in the finite harmonic crystal is described by an infinite series of the Bessel functions with multiple orders.

This phenomenon can be explained as follows.
The solution of the initial problem \eqref{cristallMotion}--\eqref{initialCondition} due to the linearity of the system can be represented as a superposition of $N$ problems for each individual particle, where only this certain particle was randomly disturbed. For each particle elastic waves propagate to the right and left from its position in the crystal and superposition of these waves for all particles describes the thermal process in the crystal. Since the crystal
is periodic (circular), the elastic waves meet each other after they have passed half-length of the crystal. All particles were disturbed instantly, therefore the waves initiated from each particle meet simultaneously, causing a sharp short-term increase of the system's kinetic temperature --- the first thermal echo. Then the waves travel further and meet again --- the second thermal echo is realized, and so on. Each thermal echo~(\ref{result}) is expressed in terms of the Bessel functions of order~$2pN$, where $p=1,2,3,...$ is the echo number.


The crystal is discrete system that posses dispersion --- the wave speed depends on the wave length. The fastest are the long waves that travel with the speed of sound $c_s=a \omega_e$ \cite{Born1954, ECM2018}, where $\omega_e$ is an elementary frequency~\eqref{cristallMotion}, $a$ is the lattice step. These waves meet after they pass half-length of the crystal, therefore the thermal echo period is
\begin{equation}\label{Period}
\tau_0=\cfrac{L}{2c_s}=\cfrac{N}{2\omega_e},
\end{equation}
where $L=Na$ is the crystal length.
Shorter waves are slower and meet later --- therefore the thermal echo has a finite width, and consequently each next thermal echo is less prominent.

The time $t=p\tau_0$ we will call the reference time for the thermal echo number $p$. At that time the long waves from the initial disturbances
have the meeting number $p$, that causes the temperature oscillations of corresponding the thermal echo.

\subsection{Thermal echo implementations}

Using formula~\eqref{result} the kinetic temperature can be represented in the form:
\ba{tempseries}{c}
     T=T_E + T_B+T_1+T_2+...+T_p+...\,;
\\
     \displaystyle T_B=\frac{\Delta T}{2}\, J_0(4\omega_et)\qq T_p=\Delta T J_{2pN}(4\omega_et);
\ea
where $T_E$ is the equilibrium temperature~\eqref{result}, $T_B$ is the basic thermal mode, and the subsequent terms $T_p$ are the thermal modes with the number ${p=1,2,3,...}$

\begin{figure}[h]
\center{\includegraphics[width=1\linewidth]{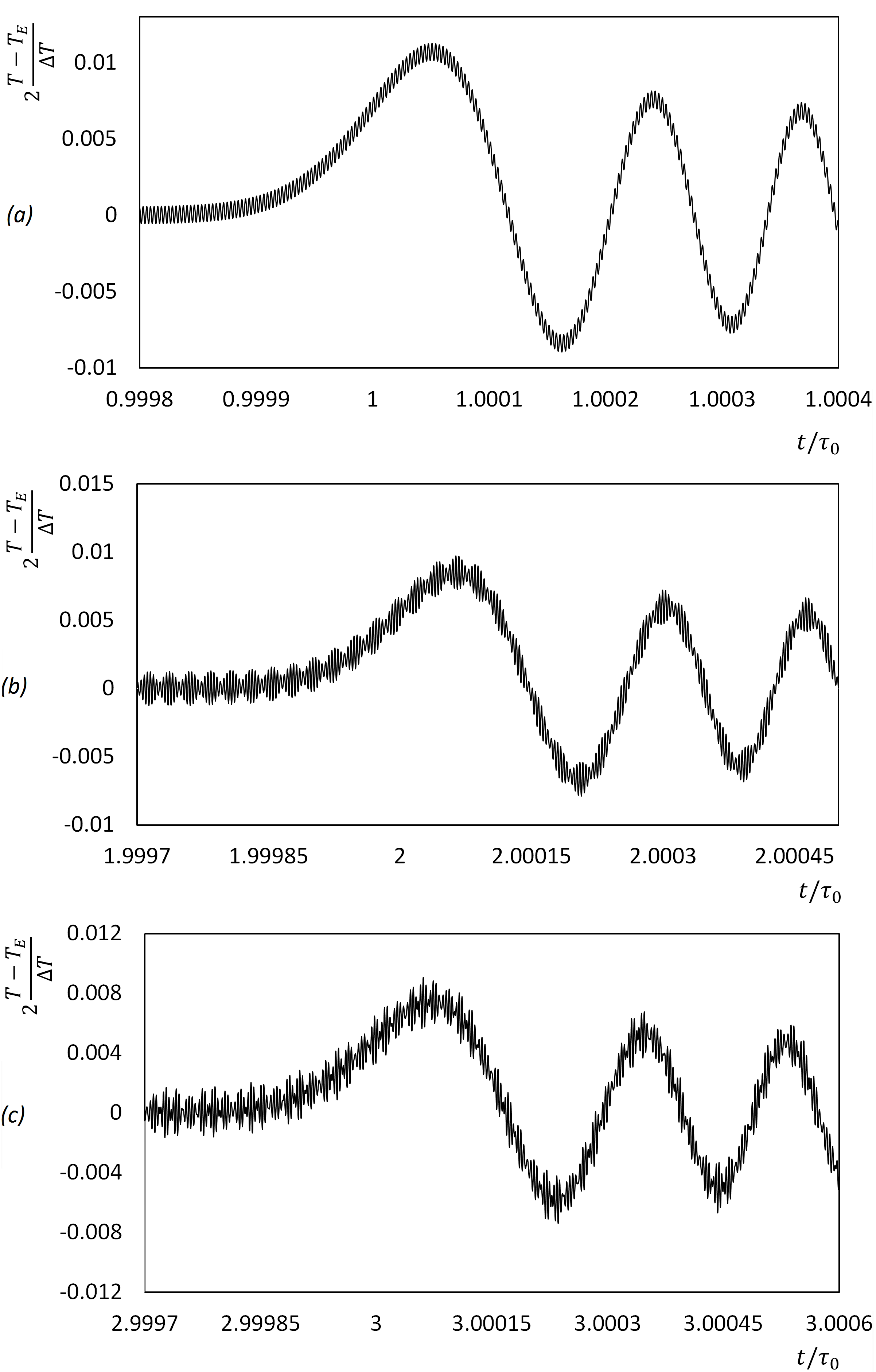}}
\caption{Beats of temperature. The number of particles $N=10^6$, $T_E$ is the equilibrium temperature, $\Delta T$ is the temperature jump of the crystal caused by the thermal perturbation, $t$ is time, $\tau_0$ is the period of realization of the thermal echo.}
\label{pic:dragon}
\end{figure}

The temperature oscillations in a crystal containing~$10^6$ particles are shown in~Fig.~\ref{pic:dragon}. The graphs show the time vicinity of the first, second and third thermal echo.
The first thermal echo is initiated for $t\approx\tau_0$.
The corresponding temperature oscillations (\ref{tempseries}) are described by the first thermal mode $T_1$.

Each thermal mode is representing by the corresponding Bessel function. Bessel functions are not periodic, however it is convenient to consider a quasiperiod --- the interval between two consequent maximums of the Bessel function. This quasiperiod is not a constant, its value decreases for each next maximum.

As can be seen from~\mbox{Fig.~\ref{pic:dragon}~(a)}, for $t\approx\tau_0$ the
oscillations are superposition~\eqref{result} of the basic mode~$T_B$ (small short oscillations on the graph) and the first mode $T_1$ (global oscillations on the graph). For $t\approx\tau_0$ the quasiperiod of the basic mode is mach smaller than the quasiperiod of the first mode. The larger is the crystal, the more significant is the discrepancy. For  times $t\approx2\tau_0$, the oscillation parameters of the basic and the first thermal modes become close, which leads to a beat phenomena.
For~$t\approx2\tau_0$, the second thermal echo is realized, superimposed by the mentioned beatings --- see~\mbox{Fig.~\ref{pic:dragon}~(b)}.
 For $t\approx3\tau_0$, the third thermal echo is realized (see~\mbox{Fig.~\ref{pic:dragon}~(c)}).
 Numerical experiments show that for large times plural realizations of the thermal echo result in an increasingly complex form of beats. The realizations of the thermal echo with large ordinal numbers is less pronounced against the background of residual oscillations from previous implementations of the thermal echo. As a result, at large times the temperature fluctuations acquire a quasi-stochastic character resembling thermal noise.

\begin{figure}[h]
{\includegraphics[width=1\linewidth]{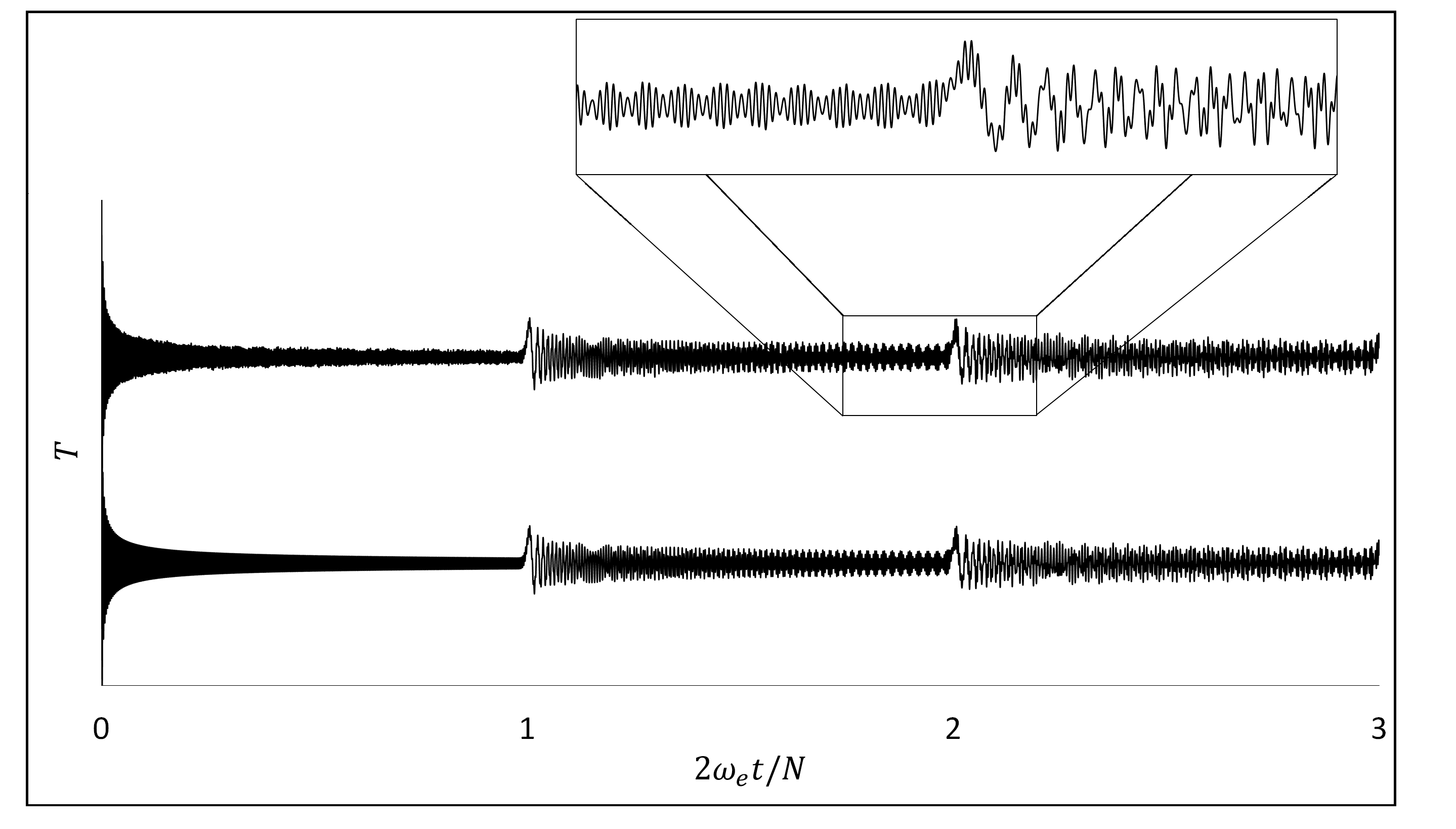}}
\caption{Oscillations of kinetic temperature $T$ in the finite crystal. Numerical (top) and analytical (bottom) solutions. The averaging is performed using 100 numerical experiments. The number of particles $N=10^3$, $t$ is time and $\omega_e$ is the elementary frequency.}
\label{ris:image1}
\end{figure}

Fig.~\ref{ris:image1} shows
comparison of the analytical solution (bottom) and the computer simulation of the crystal dynamics (top). The crystal under consideration contains $10^3$ particles. The analytical solution is described by formula~(\ref{tNsieries}) or~(\ref{result}). The computer simulation uses the method of central differences to solve numerically the system of $10^3$ differential equations of the chain dynamics~\eqref{cristallMotion} with the integration step $0.02/\omega_e$.
The results over 100 realizations of such chain with an independent random initiation are averaged.
As it can be seen from~Fig.~\ref {ris:image1} the graphs of the computer simulation and the analytical solution are almost identical.

\subsection{Asymptotics}

Expression~(\ref{result}), which describes the thermal oscillations in the crystal, includes Bessel functions of multiple orders.
In the time-vicinity of the thermal echo appearance
the following asymptotic representation of the Bessel functions~\cite{Abramowitz} can be used:
\begin{equation}\label{ass1}
  J_\mu(x)=\left(\frac{2}{\mu}\right)^{1/3}
{\rm Ai}\left(\left(\frac{2}{\mu}\right)^{1/3}(\mu-x)\right)+O(\mu^{-1}),
\end{equation}
where ${\rm Ai}$ is the Airy function~\footnote{ The Airy function can be defined as: ${\rm Ai}(x)=\frac{1}{\pi}\int_0^\infty\cos\left(\frac{t^3}{3}+xt\right){\rm d}t$} \cite{Abramowitz}. This representation is valid~\footnote{The representation \eqref{ass1} is given by the formula~$9.3.23$~\cite {Abramowitz}. It is valid for ${|\mu-x|}/{\sqrt[3]{\mu}}\leq A$, where $A$ is an arbitrary positive constant.} for ${x\approx\mu\gg 1}$.
More general asymptotics, which is valid for any $x$ is given in Appendix~\ref{BesselApp}.
The important advantage of these asymptotic representations is that they express special function $J_\mu(x)$ of two variables $x$,~$\mu$ in terms of a special Airy function of a single variable. Therefore each thermal mode~\eqref{tempseries} except the basic mode can be obtained from the Airy function by a linear transformation.
The Airy function graph in a form of ${\rm Ai}(-z)$ is depicted in Fig.~\ref {pic:airy}. This graph demonstrates the shape of the thermal echo in the thermodynamic limit~($N\to\infty$), where the value $z=0$ corresponds to the reference time $t=p\tau_0$ of the thermal echo number $p$.
It is a bit unexpected, that the significant temperature increase starts before the reference time, that is before the long waves from the initial disturbance formally meet each other. Probably this is due to the nature of discreet systems, where some energy can be transferred faster then the speed of sound of the corresponding continuum system.

Then the temperature \eqref{result} can be represented as
\begin{equation}\label{resultAiWithTe}
\begin{split}
  T \approx T_E +&\frac{\Delta T}{2}\,J_0(4\omega_et)+\\
  &\Delta T\sum_{p=1}^{\infty} \frac{1}{\sqrt[3]{pN}}\,
{\rm Ai}\!\left(\frac{2pN-4\omega_et}{\sqrt[3]{pN}}\right).
\end{split}
  \end{equation}

For large values of arguments $x\gg\mu+1$ the following asymptotics for the Bessel functions fulfils~{\cite{Abramowitz}:
 \begin{equation}\label{ass2}
  \begin{split}
 J_\mu(x)=\sqrt{\dfrac{2}{\pi x}}\,\cos\left(x-\dfrac{\pi \mu }{2}-\dfrac{\pi}{4}\right)+O(x^{-2}).
  \end{split}
 \end{equation}

Asymptotic representations \eqref{ass1} and~\eqref{ass2} for the Bessel functions allows to obtain the basic characteristics of the temperature oscillations for large~$N$.

\begin{figure}[h]
{\includegraphics[width=0.95\linewidth]{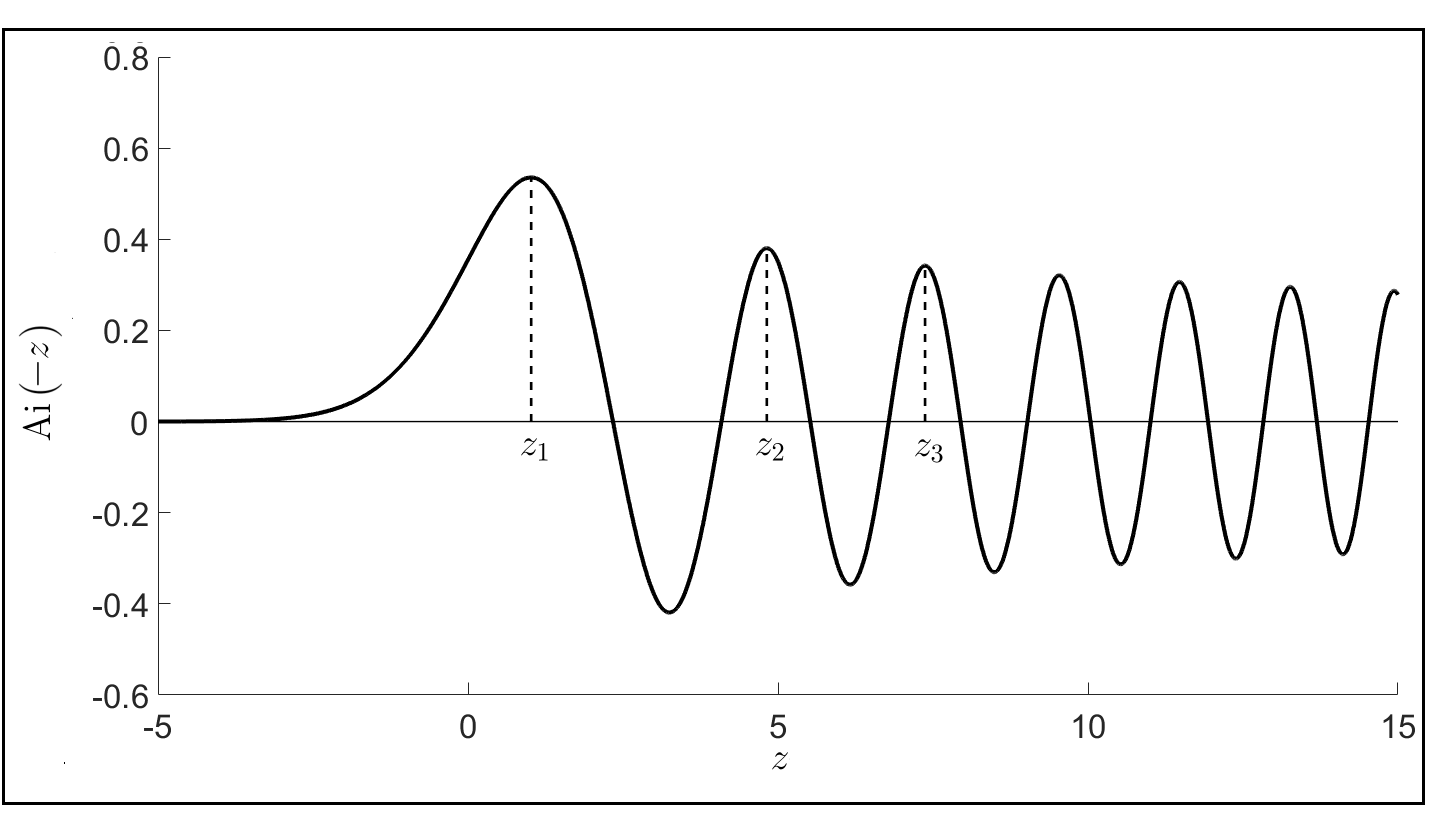}}
\caption{The Airy function. Such a form takes any thermal echo for a sufficiently large number of particles~$N$.}
\label{pic:airy}
\end{figure}

\subsection{Characteristics of thermal echo}

Let us consider the thermal mode number $p$:
\be{Tp}
      T_p  = \nu_p \Delta T \,J_{2pN}(4\omega_et)
\ee
for large values of $N$, where $\nu_p=\frac{1}{2}$ for $p=0$ (the basic thermal mode) and~$\nu_p=1$ otherwise.

In the vicinity of the reference time $t=p\tau_0$ the thermal mode $T_p$
with the use of asymptotics~(\ref{ass1}) can be represented as
\be{Tp1}
      T_p\simeq\frac{\nu_p\Delta T}{\sqrt[3]{pN}}\,{\rm Ai}\!\left(\frac{2pN-4\omega_et}{\sqrt[3]{pN}}\right).
\ee
The Airy function ${\rm Ai}(-z)$ is depicted in Fig.~\ref{pic:airy}. Let $z_k$ be the successive points of the local maximums of this function, where $k=1,2,3,...$ is the point number; $A_k\={\rm Ai}(-z_k)$ are the corresponding maximums of the function. The first three values of these constants are~\cite{Fabijonas1999}:
\begin{equation}\label{maxAi}
\begin{split}
   &z_1\approx1.0:\quad A_1\approx0.53, \\
   &z_2\approx4.8:\quad A_2\approx0.38, \\
   &z_3\approx7.4:\quad A_3\approx0.34.
\end{split}
\end{equation}
Then the corresponding points of the first local maximums $t_p$ and the maximum values $M_p$ for the thermal mode (\ref{Tp1}) are
\be{maxJ}
  t_p\simeq\frac{1}{4\omega_e}\Big(2pN+\sqrt[3]{p N}\Big)
\qq
  M_p\simeq\frac{\nu_p\Delta T}{\sqrt[3]{p N}}A_1
.\ee
The formula for $t_p$ can be rewritten in the form
\be{maxJ-1}
  t_p\simeq p\tau_0\left(1+\frac{1}{2(p N)^{2/3}}\right)
.\ee
Hence the relative difference between the reference time and the time of the temperature maximum $t_p$ decreases with increase of $N$, which proves that these times coincide in the continuum limit.

In the vicinity of the reference time $t=p\tau_0$ the previous thermal mode $T_{p-1}$~(\ref{Tp}) with the use of asymptotics~(\ref{ass2}) can be represented as
\be{Tp2}
    T_{p-1} \simeq
    (-1)^N\nu_{p-1}\frac{\Delta T}{\sqrt{\pi N p}}\,\cos\left(4\omega_et-\pi N p-\dfrac{\pi}{4}\right).
\ee


Let us define the relative height of the thermal echo $h_p$ as a ratio of the maximum value of the thermal mode number $p$ (\ref{maxJ}) to the amplitude of the residual oscillations from the previous thermal mode (\ref{Tp2}), which gives~\footnote{For $p=1$ the previous mode is $T_B$ with twice less amplitude --- see (\ref{tempseries}).}
\begin{equation}\label{defrelHigh}
  h_p \simeq \frac{\sqrt\pi A_1\sqrt[6]{p N}}{\nu_{p-1}}
.\end{equation}
Thus the relative height of the thermal echo~$h_p$ increases with increasing~$N$.
Therefore for very large $N$ the residual oscillations can be considered as negligible. However this increase is very slow --- proportional to $N^{1/6}$, hence even for $N=10^6$ the residual oscillations are quite noticeable.


To analyze the thermal echo properties let us introduce its relative width~$w_k$ as the ratio of the quasiperiod of the corresponding Bessel function to the asymptotic period of Bessel functions at infinity. From formulas (\ref{ass1})--(\ref{ass2}) it follows that the relative width is proportional to $N^{1/3}$ (see appendix~\ref{app3}):
\begin{equation}\label{deltax}
  w_k\simeq\frac{z_{k+1}-z_{k}}{2\pi}\,\sqrt[3]{pN}.
\end{equation}

From relations (\ref{defrelHigh}) and (\ref{deltax}) it follows that for large $N$ the residual oscillations from the previous thermal echo are small and frequent in comparison with the current thermal echo wave --- see Fig.~\ref{pic:dragon}.


\subsection{Example}
As an example let us consider a chain of $N=10^3$ carbon atoms forming a ring with the initial temperature~\mbox{$T_0=0^\circ$C}. The heat perturbation
instantaneously increases the temperature up to the value
\begin{equation}
T_0+\Delta T=100^\circ\mbox{C}.
\end{equation}
The chosen value of $\Delta T$ is sufficiently small in comparison with the melting temperature of carbon~($3500^\circ$C), so as not to take into account the nonlinearity of the interatomic interaction.
As a result of this perturbation the temperature oscillations near the equilibrium value of \mbox{$T_E\approx50^\circ$C} are realized in the crystal. The graph of the temperature oscillations for the crystal calculated using numeric and analytical solution is shown in
~Fig.~\ref{ris:image1}.
Formula \eqref{maxJ} gives for the maximum value of the first thermal mode
\begin{equation}\label{maxEcho}
    M_1\approx 5.3^\circ {\rm C}
\qquad\Longrightarrow\qquad
    T_E+ M_1\approx 55.3^\circ \rm C
.\end{equation}
Thus the thermal echo brings approximatly $10\%$ change of the temperature comparing to the final temperature raise $T_E-T_0 \approx 50^\circ \rm C$.

These results are slightly perturbed by the residual oscillations from the  previous (the basic) thermal mode.
Formulas~(\ref{defrelHigh}) and~(\ref{deltax}) yield
\begin{equation}\label{delta}
h_1\approx{5.9}\qq w_1\approx11.3
,\end{equation}
hence the residual oscillations are approximately $6$ times weaker and have~$11$ times shorter period than the oscillations caused by the first thermal echo.
The residual oscillations according to formula \eqref{Tp2} have the amplitude of approximately $0.9^\circ \rm C$, which is about~$17\%$ of~$M_1$.
If the residual oscillations are taken into account the maximum temperature achieved by the first thermal echo is approximately $56.2^\circ \rm C$.

The above characteristics depend only on the number of particles $N$, while the period $\tau_0$ of the thermal echo realization depends on the physical properties of the crystal and the type of oscillations (longitudinal or tranversal). Let us consider longitudinal oscillations, the mass of the carbon atom \mbox{$m=1.99\cdot10^{-26}$}~kg and the stiffness of diamond  bond ${C=1824}$~N/m \cite{Berinskii2016}. Then formula~(\ref{Period}) gives \mbox{$\tau_0=1.65\cdot10^{-12}$~s} that is a $\frac{N}{4\pi}\approx80$ times greater than the atomic oscillation period.

The obtained numeric characteristics of the thermal echo can be used to determine its occurrence in natural experiments.

\section{Conclusions}

The paper considers a finite one-dimensional harmonic crystal subjected to an instant spatially uniform thermal perturbation.
The numeric and analytical analysis presented in the paper demonstrates the phenomenon of the thermal echo: a sharp short-term temperature rise that is periodically realized in the crystals.

Previous papers~\cite{Krivtsov 2014 DAN, Prig} have shown that in the infinite harmonic crystal the instant thermal perturbation produces the thermal oscillations with a monotonically decreasing amplitude, these oscillations are described by the zero order Bessel function. In the present paper it is shown that in the finite harmonic crystal the sequence of realizations of the thermal echo is described by a series of the Bessel functions of multiple orders. Any thermal echo in the thermodynamic limit is described by the Airy function. A superposition of the temperature oscillations generated by the sequential thermal echoes results in a temperature beats. Each subsequent thermal echo complicates the shape of the beats.


It follows thus from the analysis that the maximum temperature increase caused by the thermal echo decreases as~$\sqrt[3]{pN}$~\eqref{maxJ}, where $p$ the thermal echo number and $N$ is the number of particles in the crystal. The width of the thermal echo grows by the same law~\eqref{deltax}.
Between any two 
thermal echoes the amplitude of the temperature oscillations decreases in proportion to the square root of time~\eqref{ass2}. The larger is the crystal, the more noticeable are the temperature increases against the residual oscillations~\eqref{defrelHigh}.

Thus, an analytical description of the phenomenon of the thermal echo is presented. This phenomenon is an important feature of thermal processes in finite systems and should be taken into account in the development of the modern micro- and nanosize electronic devices.

\begin{acknowledgments}
The authors would like to express their sincere gratitude to W.G. Hoover, D.A. Indeitsev and V.A. Kuzkin for their time and useful discussions.
\end{acknowledgments}

\vspace{\baselineskip}

This work was supported by the Russian Science Foundation (grant 18-11-00201).

\appendix
\section*{Appendixes}
\addcontentsline{toc}{section}{Appendixes}
\appendix
\section{The initial problem for the velocities covariance}\label{app00}
Let us consider displacements covariance and velocities covariance:
\begin{equation}\label{disvelcov}
    \xi_n=\av{u_ku_{k+n}}\qq \kappa_n=\av{v_kv_{k+n}},
\end{equation}
where ${\cal L}$ the operator from the problem \eqref{cristallMotion}.
Differentiation of expression~\eqref{disvelcov} gives us the equation
\begin{equation}
    \dot\kappa_n={\cal L}\dot\xi_n,
\end{equation}
that leads us to the conservation law:
\begin{equation}\label{conservationLaw}
    \kappa_n-{\cal L}\xi_n=\epsilon_n^0,
\end{equation}
where the constant $\epsilon_n^0$ can be obtained from the initial conditions~\eqref{initialCondition}:
\begin{equation}\label{epsilon}
    \epsilon_n^0=\kappa_n\bigg|_{t=0}=\sigma^2\av{\tilde\rho_k\tilde\rho_{k+n}}\bigg|_{t=0},
\end{equation}
where ${\tilde\rho_k}$ is centered random numbers. Let us consider the following expression:
\begin{equation}\label{matexp}
    \av{\tilde\rho_k\tilde\rho_{k+n}}=\av{\rho_k\rho_{k+n}}-2\av{\rho_k\overline{\rho}}+\av{\overline{\rho}^2},
\end{equation}
where $\overline{\rho}$ is the mean value of $\rho_k$. Let us take into account that
\begin{equation}\label{tiszero}
    t=0:\quad \av{\rho_k\overline{\rho}}=\frac{\rho_k}{N}\sum_{n=0}^{N-1}\rho_{k+n}=\frac{1}{N}\qq \overline{\rho}^2=\frac{1}{N},
\end{equation}
then
\begin{equation}\label{epsilongen}
    \epsilon_n^0=\sigma^2\delta_n-\frac{\sigma^2}{N}.
\end{equation}
The equations below can be obtained by differentiating twice~\eqref{disvelcov}:
\begin{equation}\label{dd}
\begin{split}
    &\ddot\xi_n=2(\kappa_n+{\cal L}\xi_n) ,\qquad
    \ddot\kappa_n=2{\cal L}(\kappa_n+{\cal L}\xi_n).
\end{split}
\end{equation}
The use of conservation law~\eqref{conservationLaw} gives us a closed equation for each covariance:
\begin{equation}\label{dd}
\begin{split}
    &\ddot\xi_n-4{\cal L}\xi_n=2{\cal L}\epsilon_n^0, \qquad
    \ddot\kappa_n-4{\cal L}\kappa_n=-2{\cal L}\epsilon_n^0.
\end{split}
\end{equation}
The second equation of \eqref{dd} and \eqref{epsilon} gives us the initial problem for $\kappa_n$.
\section{Equation for velocities covariances} \label{app1}

Let us consider the initial problem \eqref{DDOTkappa} with the corresponding boundary conditions:
\begin{equation}\label{k1}
\begin{split}
      &\ddot\kappa_n-4{\cal L}\kappa_n=-2{\sigma^2}{\cal L}\delta_n, \\
     & t=0: \quad \kappa_n={\sigma^2}\delta_n-\frac{\sigma^2}{N}\qq \dot \kappa_n=0,\\
  &\kappa_0= \kappa_{N}\qq \kappa_{N+1} = \kappa_1.
\end{split}
\end{equation}
Then we apply the discrete Fourier transformation
\begin{equation}\label{k20}
    {\kappa}^*_k=\sum_{n=0}^{N-1}\kappa_ne^{\frac{-2\pi i k n}{N}}
\end{equation}
to the initial problem~\eqref{k1}:
\begin{equation}\label{k2}
\begin{split}
  &\ddot{\check{\kappa}}_k+4{\rm L}\check{\kappa}_k=0, \\
    & t=0: \quad \check{\kappa}_k=\frac{\sigma^2}{2}-\sigma^2\delta_k\qq \dot{\check{\kappa}}_k=0,\\
\end{split}
\end{equation}
where
\begin{equation}\label{k21}
     \check{\kappa}_k={\kappa}^*_k-\frac{\sigma^2}{2}\qq {\rm L}=4\omega_e^2\sin^2\frac{\pi k}{N}.\\
\end{equation}
The solution of the problem \eqref{k2} is
\begin{equation}\label{k3}
    \check{\kappa}_k=\frac{\sigma^2}{2}\left(1-2\delta_k\right)\cos\left(4\omega_e t\sin\frac{\pi k}{N}\right),\\
\end{equation}
hence
\begin{equation}
   {\kappa}^*_k=\frac{\sigma^2}{2}\left[\left(1-2\delta_k\right)\cos\left(4\omega_et\sin\frac{\pi k}{N}\right)+1\right].
\end{equation}
The inverse Fourier transform for $\kappa_k^*$ is
\begin{equation}\label{k32}
\begin{split}
    &\kappa_n=\\
    &\frac{1}{N}\frac{\sigma^2}{2}\sum_{k=0}^{N-1}\left(1+\cos\left(4\omega_e t\sin\frac{\pi k}{N}\right)\right)e^{i\frac{2\pi nk}{N}}-\frac{\sigma^2}{N}.
\end{split}
\end{equation}
The $\kappa_n|_{n=0}$ which is proportional $\Delta T$ is
\begin{equation}\label{k4}
\begin{split}
    &\kappa_n|_{n=0}=\frac{\sigma^2}{2}\left(1+\frac{1}{N}\sum_{k=1}^{N-1}\cos\left(4\omega_et\sin\frac{\pi k}{N}\right)\right).
\end{split}
\end{equation}

\section{The formula for the sum of cosines of multiple angles} \label{app2}

Let us demonstrate that the sum of the multiple angles of the cosine can be represented by the following form:
\eq{sumcos1}{
\frac{1}{N}\sum_{k=0}^{N-1}\cos\left(p \frac{2 \pi k}{N}\right)=
\frac{\sin (2\pi p)}{2N}\,{\rm ctg \frac{\pi p}{N}}+\frac{\sin^2(\pi p)}{N}.
}
The sum of cosines of multiple angles can be calculated as a sum of exponentials, which is calculated as the sum of the geometric progression:
\begin{equation}\label{geom}
  \sum^{N-1}_{k=1}\cos k\phi={\rm Re}\sum^{N-1}_{k=1}e^{ik\phi}={\rm Re}\left(\frac{e^{iN\phi}-e^{i\phi}}{e^{i\phi}-1}\right).
\end{equation}
Calculation of the real part (\ref{geom}) gives us one of the following representations:
\begin{equation}
\begin{split}
   &\sum^{N-1}_{k=0}\cos k\phi=\\
   &\frac{\sin(N\phi-\phi/2)}{2\sin({\phi}/{2})}-\frac{1}{2}=\frac{\cos({N\phi}/{2})
\sin((N-1)\phi/2)}{\sin({\phi}/{2})}=\\
   &=\frac{1}{2}\sin(N\phi){\rm ctg}(\phi/2)-\cos^2(N\phi/2).
\end{split}
\end{equation}
For sum
\begin{equation}\label{defSNP}
  F_N(p)\= \frac{1}{N}\sum_{k=1}^{N-1}\cos\left(p\frac{2\pi k}{N}\right)
\end{equation}
we get the following representation:
\begin{equation}\label{prapra}
      F_N(p)=\frac{\sin (2\pi p)}{2N}\,{\rm ctg \frac{\pi p}{N}}+\frac{\cos^2(\pi p)}{N}.
\end{equation}

For all $p$, except $p$ that divides by $N$, the expression $F_N(p)$ from (\ref{prapra}) is equal to $1-1/N$, and for $p$ divides by $N$, the value of the expression $F_N(p)$ is found from~(\ref{defSNP}), and is equal to $-1/N$.

\section{Asymptotics for the Bessel functions}\label{BesselApp}
An asymptotic for the Bessel functions of large orders $\mu$ and arbitrary argument $x$ is represented by the series $9.3.35$ from \cite {Abramowitz}.
The first term of this expression can be represented as the following composition:
\begin{equation}\label{represGen}
    J_{\mu}(x)\simeq\,\left(\frac{4\zeta}{1-\xi^2}\right)^{1/4}\,\frac{{\rm Ai}\left(\mu^{2/3}\zeta\right)}{\mu^{1/3}}\qq
 \mu\to\infty,
 \end{equation}
 where
\begin{gather*}
\begin{split}
&\xi=\frac{x}{\mu}\\
&\zeta=
  \begin{cases}
   &\left(\cfrac{3}{2}\right)^{2/3}\,\left[\ln\left(\cfrac{1+\sqrt{1-\xi^2}}{\xi}\right)-\sqrt{1-\xi^2}\right]^{2/3},\\[3ex] 
   &\text{if } {0\leq \xi\leq1};\\[3ex]
   &-\left(\cfrac{3}{2}\right)^{2/3}\,\left[\sqrt{\xi^2-1}-{\rm arcsec}\,{\xi}\,\right]^{2/3},\\[3ex] 
   &\text{if } {\xi\geq1},
 \end{cases}
\end{split}
   \end{gather*}
and $\rm Ai$ is the Airy function \cite{Abramowitz}.

In the case of the $\xi\to1$, the representation \eqref{represGen} is reduced to the following one:
\begin{equation}\label{approxxi1}
      J_\mu(x)\approx\left(\frac{2}{\mu}\right)^{1/3}
{\rm Ai}\left(2^{1/3}\mu^{2/3}(1-\xi)\right),
\end{equation}
 where the approximation \mbox{$\zeta\approx2^{1/3}(\xi-1)$} is used.
 The~\eqref{approxxi1} is equivalent to representation~\eqref{ass1}, considered before.

\section{Relative width of the thermal echo}\label{app3}
Let us denote by $\{x_i\}$ the set of points in which the Bessel function $J_\mu(x)$ has local maxima ($x_1$ is the first local maximum, $x_2$ is the second, etc.). The approximate values of $x_i$ are found from the condition that the argument of the Airy function in the expression (\ref{ass1}) is equal to the values $-z_i$ \eqref{maxAi}:
 \begin{equation}\label{maxJin}
\begin{split}
    &x_i\approx\mu+z_i\left(\frac{\mu}{2}\right)^{1/3}\qq J_{\mu}(x_i)\approx\left(\frac{\mu}{2}\right)^{-1/3}{\rm Ai}(-z_i).\\
\end{split}
 \end{equation}
The value of the asymptotic period of the Bessel function is~$2\pi$ as seen from the formula~\eqref{ass2}. Then the expression for the width of the thermal echo becomes:
 \begin{equation}
 w_i\stackrel{\mbox{\scriptsize def}}{=}\frac{x_{i+1}-x_i}{2\pi}\approx\frac{z_{i+1}-z_{i}}{2\pi}\left(\frac{\mu}{2}\right)^{1/3}.
\end{equation}

\end{document}